# EVAPORATION RATE OF WATER IN HYDROPHOBIC CONFINEMENT

by

Sumit Sharma[a] and Pablo G. Debenedetti[a,1]

[a]Department of Chemical and Biological Engineering, Princeton University, Princeton, NJ 08544

[1] To whom correspondence should be addressed. Email: pdebene@princeton.edu

**The drying of hydrophobic cavities is believed to play an important role in biophysical phenomena such as the folding of globular proteins, the opening and closing of ligand-gated ion channels, and ligand binding to hydrophobic pockets. We use forward flux sampling, a molecular simulation technique, to compute the rate of capillary evaporation of water confined between two hydrophobic surfaces separated by nanoscopic gaps, as a function of gap, surface size and temperature. Over the range of conditions investigated (gaps between 9 and 14 Å and surface areas between 1 and 9 nm$^2$) the free energy barrier to evaporation scales linearly with the gap between hydrophobic surfaces, suggesting that line tension makes the predominant contribution to the free energy barrier. The exponential dependence of the evaporation rate on the gap between confining surfaces causes a ten order-of-magnitude decrease in the rate when the gap increases from 9 to 14 Å. The computed free energy barriers are of the order of 50$kT$, and are predominantly enthalpic. Evaporation rates per unit area are found to be two orders of magnitude faster in confinement by the larger (9 nm$^2$) than by the smaller (1nm$^2$) surfaces considered here, at otherwise identical conditions. We show that this is a direct consequence of the dependence of hydrophobic hydration on the size of solvated objects. For sufficiently large surfaces, the critical nucleus for the evaporation process is a gap-spanning cylindrical vapor tube.**

**Introduction**

The behavior of water near hydrophobic surfaces is of interest in a wide range of technological contexts. Examples include the design of self-cleaning materials (1) and anti-ice coatings (2), and the development of novel processes for the storage and dissipation of mechanical energy (3). Scientifically, many aspects of hydrophobic hydration are the object of active inquiry (4); examples include the role of density fluctuations in nanoscopic hydrophobic interfaces (5), the entropic or enthalpic character of hydrophobic hydration and its dependence on solute size and thermodynamic conditions (6, 7), and the molecular conformations and solubility of long-chain alkanes in water (8). A fundamental connection between hydrophobicity and biological self-assembly was first pointed out by Walter Kauzmann (9), who showed that the water-mediated tendency for apolar moieties to aggregate is crucial for protein conformational stability. Tanford's work further contributed to establishing the centrality of water-mediated interactions in biological self-assembly (10). Since these seminal insights, the view has gradually emerged of water as an active participant in life's processes (11).

Water confined by two impenetrable surfaces is the simplest example of water-mediated interactions between (large) hydrophobic objects. When the distance between such hydrophobic surfaces falls below a critical value, evaporation of water is favored thermodynamically (12). The resulting surface-induced evaporation has been the subject of numerous theoretical and computational studies (e.g., 6, 13-19), with several focusing on biological hydrophobic interfaces (e.g., 20-22).

Previous computational studies of capillary evaporation in hydrophobic confinement have addressed the underlying thermodynamics or have been limited to phenomenological observations of the occurrence or absence of capillary evaporation in finite-time molecular dynamics simulations. Much less attention has been devoted to the equally important matter of evaporation kinetics. Notable exceptions include the important work of Luzar and coworkers (23-25), Bolhuis and Chandler (26), and Xu and Molinero (27). Leung et al. (23) used a combination of umbrella sampling and reactive flux formalism to compute the rate of capillary evaporation of SPC water (28) in a semi-infinite hydrophobic slit. Subsequently, Luzar (24) used a lattice model to investigate the dependence of the free energy barrier on the separation between the confining surfaces. Bolhius and Chandler (26) used transition path sampling to study the cavitation of the Lennard-Jones liquid between repulsive surfaces. They focused on the nature of the transition state, and pointed out the relevance of their findings to the hydrophobic effect. Xu and Molinero (27) studied the thermodynamics and kinetics of liquid-vapor oscillations in a coarse-grained model of water in nano-scale hydrophobic confinement.

The drying of hydrophobic cavities is thought to be important in biophysical phenomena such as the folding of globular proteins (4, 6, 9, 10, 20), the opening and closing of ligand-gated ion channels (29), and ligand binding to hydrophobic pockets (30). Thus, knowledge of the rate of capillary evaporation in hydrophobic confinement, and its dependence on temperature, pressure, confinement length scale, size of the confining surfaces, and surface characteristics such as degree of hydrophobicity and curvature, should be useful for a quantitative understanding of several important biophysical phenomena. In this paper we report on a computational investigation of the effects of surface size, confinement length scale and temperature on the kinetics of capillary evaporation of water in hydrophobic confinement.

As will be shown, evaporation requires the formation of a sufficiently large void in the confined region, a rare event. A straightforward molecular dynamics (MD) simulation is therefore incapable of providing quantitative rate information on the basic phenomenon of interest here. Accordingly, we use Forward Flux Sampling (FFS), a technique specifically designed to sample rare events (31-33), in conjunction with MD. Figure 1 shows schematically the implementation of the calculation. We consider two hydrophobic surfaces separated by a gap $d$ immersed in water at fixed temperature and pressure, and use the FFS technique to calculate the rate at which the confined volume $L^2d$ is emptied. We perform the calculation for a range of values of $d$, $L$, and temperature. Technical details are provided in the Methods section.

## Results and Discussion
**Rates.** Figures 2 and 3 show the calculated evaporation rate as a function of the gap between hydrophobic surfaces, $d$, for 1.0 x 0.9 (Fig.2) and 3.2 x 3 nm$^2$ surfaces (Fig. 3), at 298 K and 1 bar (henceforth we refer to these as 1 x 1 and 3 x 3 nm$^2$ surfaces, respectively). The characteristic time $\tau$ required to nucleate a surface-induced evaporation event is given by [$\tau \sim (jA)^{-1}$], where $A$ is the surface area and $j$ is the evaporation rate. This time increases by 10 orders of magnitude (from 6.3 x 10$^{-10}$ sec to 17.2 sec) as the gap between small (1 x 1 nm$^2$) surfaces increases from 9 to 14 Å; similarly, there is a 6-order-of-magnitude increase in the characteristic evaporation time between large (3 x 3 nm$^2$) surfaces upon increasing the gap from 11 to 14 Å. These numbers suggest constraints on the range of gaps for which capillary evaporation can occur at rates that are dynamically relevant to biophysical phenomena.

**Free energy barrier and gap dependence.** In general, the evaporation rate $j$ can be expressed as

$$j = C\exp[-\Delta G(d)/kT] = C'\exp[-\Delta H(d)/kT] \tag{1}$$

where $C$ is a gap-independent pre-exponential factor, $\Delta G$ is the free energy barrier to nucleation, $C' = C \exp[\Delta S(d)/k]$, $\Delta S$ and $\Delta H$ are the entropic and enthalpic contributions to $\Delta G$, and $k$ is Boltzmann's constant. Equation (1) implies that by computing the evaporation rate as a function of the gap $d$ and temperature, one can extract information on $\Delta G$, $\Delta H$ and $\Delta S$. The numerical procedures used to fit the rate data to Equation (1) and to regress values for $\Delta G$, $\Delta H$ and $\Delta S$ are described in the Supporting Information. Briefly, from an Arrhenius plot, ln$j$ vs. $1/T$ (Figure 2, inset) we obtain both ln $C'$ (intercept) and $\Delta H$ (slope = -$\Delta H/k$). ln $C'$ was found to be independent of $d$, implying that the entropic contribution to the free energy is either small or $d$-independent. Using the last expression in Equation (1), $\Delta H$ was found to scale linearly with $d$, which implies that $\Delta G$ is also linear in $d$. With $\Delta G = A + Bd$, and hence ln$j$ = ln $C$ – $A/kT$ – $Bd/kT$ we obtain $A$, $B$ and ln $C$ by regression of the computed rates, $j$ ($T$, $d$). Finally, $\Delta S$ is given by the ratio of the intercepts, $\Delta S/k$ = ln($C'/C$). We find that the free energy barrier is predominantly enthalpic, with $T \Delta S/\Delta H \sim O(10^{-3})$ and $O(10^{-1})$ for the small and large surfaces, respectively. Over the range of conditions investigated in this work, we find that the rate of change of the free energy barrier with respect to the gap, $B$, is between 4 and 5 $kT$/Å. Table 1 compares the free energy barriers computed directly from Equation (1) with the values obtained by rescaling $\Delta G$ (at 9.8 Å for 1 x 1 nm$^2$ surfaces; at 12 Å for 3 x 3 nm$^2$ surfaces) assuming linear scaling, $\Delta G \sim d$. The good agreement shows that, over the range of conditions explored in this work, the free energy barrier scales linearly with the gap between hydrophobic surfaces. As documented in the Supporting Information, neither a quadratic dependence, $\Delta G \sim d^2$, nor using ($d – 2l$) instead of $d$ to fit the data, yielded accurate representations of the evaporation rate (here, $l$ is the thickness of the vapor layer adjacent to the hydrophobic surface, which can be clearly seen in Figure 1; see Supporting Information for details on the determination of $l$). It is important to note that in this work we use an indirect, kinetic route to calculate *ΔG*. It would be useful to compute this quantity directly, using free energy sampling techniques.

The linear scaling of *ΔG* with *d* suggests that the predominant contribution to *ΔG*, over the range of conditions, size of the hydrophobic surfaces, and gaps investigated here, comes from line tension (35, 36). To rationalize this, we consider the formation of a cylindrical vapor tube of radius *r* between two solid surfaces separated by a distance *d*, the gap being otherwise filled with liquid (Figure S1). As will be shown below, the critical nucleus for sufficiently large surfaces is indeed a cylindrical tube. The equilibrium state of a macroscopic system corresponds to a condition of minimum free energy (e.g., minimum Gibbs free energy for a closed system at fixed temperature and pressure; minimum Helmholtz free energy for a closed system at fixed temperature and volume). For an open system possessing both an interface (e.g., vapor-liquid) and a line along which three phases are in contact (e.g., solid-liquid-vapor), this free energy is given by $\Omega = -PV + \gamma F + \lambda L$, and is called the grand potential. Here, *P* denotes pressure; *V*, volume; *F*, interfacial area; *L*, the linear dimension along which three phases are in contact; *γ*, the vapor-liquid interfacial tension; and *λ* is the line tension associated with three-phase contact along the circumference of the cylinder's base. The free energy cost of forming a gap-spanning vapor tube is given by

$$\Delta\Omega = \pi r (d\gamma + 4\lambda) - 2\pi r^2 \gamma \tag{2}$$

The above expression assumes that the surface is perfectly non-wetting (contact angle 180°). The derivation of Equation (2) is provided in the Supporting Information.

The free energy maximum occurs for a tube radius *r*\*, given by

$$r^* = \frac{d}{4} + \frac{\lambda}{\gamma} \tag{3}$$

in correspondence to which the free energy barrier is

$$\Delta\Omega = \frac{\pi\gamma d^2}{8} + \pi\lambda d + \frac{2\pi\lambda^2}{\gamma} \tag{4}$$

In the absence of a line tension contribution, the free energy barrier scales quadratically with the gap, a well-known result (14). Using typical values [$\gamma \sim 0.07$ N/m (12), $d \sim 1$nm, $\lambda \sim 10^{-5}$dyn (35, 37)], the relative magnitude of the three terms on the right hand side of Equation (4) is $\sim$ 1: 11: 33, indicating that line tension makes the predominant contribution to the free energy barrier. The literature includes reports of positive as well as negative line tensions (37). Our observations are consistent with positive line tensions of magnitudes such as are reported in the literature (35, 37).

**Surface size dependence.** As shown in Figures 2 and 3, evaporation rate calculations were performed at 11, 12, 13 and 14 Å gaps for both the small (1 x 1 nm²) and large (3 x 3 nm²) surfaces. For a given gap, evaporation is much faster for the larger surfaces: the rate for the 3 x 3 nm² hydrophobic surfaces is 40 times larger than for the 1 x 1 nm² surfaces when the gap is 11 Å, and 358 times larger when the gap is 14 Å. Table 2 lists the average water density and compressibility in the confined region for the different gap sizes, and for small (1 x 1 nm²) and large (3 x 3 nm²) surfaces. Effective compressibilities were obtained from the fluctuation equation $K_T = V<(\delta\rho)^2>/<\rho>^2 kT$, where $K_T$ is the isothermal compressibility, $V$ is the confined volume, and angle brackets denote thermal average.

It can be seen that, for a given value of the gap, the density of confined water decreases and its compressibility increases, as the size of the confining surface increases. This observation is consistent with Stillinger's important insight regarding the structure of aqueous interfaces near large non-polar objects (38), with the theoretical description of the manner in which soft interfaces arise on nanoscopic scales (6), and with subsequent results from simulations of capillary evaporation using lattice models (25). Thus, the marked increase in evaporation rate with the size of the confining surfaces is a manifestation of the length-scale dependence of hydrophobicity, whereby the interface between water and a hydrophobic object evolves from hard and liquid-like to soft and vapor-like as the size of the solvated object increases (6, 7). Accordingly, penetration into the metastable region is accomplished both by bringing a given pair of hydrophobic surfaces closer together or by enlarging the hydrophobic surface area while keeping the gap unchanged.

A complementary theoretical perspective on the size-dependence of evaporation for a given gap follows from considering the critical gap $d_c$, between hydrophobic surfaces below which confined liquid water becomes metastable with respect to the vapor. This quantity is given by (6, 12, 13, 24)

$$d_c = \frac{2\gamma}{\Delta p \left[1 + \frac{4\gamma}{L \Delta p}\right]} \tag{5}$$

where $\Delta p$ is the difference between the imposed pressure and the saturation pressure at the given temperature, and the immersed surfaces are assumed to be $L$ x $L$ squares. For $L \sim 1$ nm, the second term in brackets in the right hand side denominator of Equation (5) is of order $10^3$, whereupon the following simplified result follows (12):

$$d_c \approx L/2 \tag{6}$$

This implies that if nano-scale pairs of hydrophobic surfaces of different size are immersed in water, the supersaturation will increase with the size of immersed surface, even if the gap between pairs of surfaces is fixed. Hence we expect that the evaporation rate will increase with the characteristic size of the hydrophobic surfaces. Generalization of Equations (5) and (6) to include line tension is discussed in the Supporting Information.

Equation (6) also suggests that for the 1 x 1 nm² surfaces, the vapor may be metastable with respect to the confined liquid since $d > L/2$. It should be emphasized, however, that the continuum picture on which Equation (6) is predicated breaks down at molecular length scales (12). Thus, it is the scaling $d_c \sim L$, not the precise coefficient, that is sufficient to rationalize the $L$-dependence of the computed evaporation rates.

**Transition state.** In order to investigate the nature of the transition state leading to evaporation, calculations were performed at gaps of 9.8 Å (small surfaces) and 12 Å (large surfaces), at 298K and 1 bar. Configurations that upon randomizing the molecular velocities have equal probability of reaching the vapor state (empty gap space) or returning to the liquid state constitute the transition state ensemble (26, 39-41). Members of this ensemble were harvested by a three-step computational procedure described in Methods.

The fraction of trajectories that, starting from a given configuration, reach the vapor state without first returning to the liquid state constitutes the committor probability for that configuration (41). Figures 4 and 5 show the committor probabilities for the various configurations. Each curve corresponds to a fixed number of water molecules in the confined region, $N$ (small surfaces) or to a range of $N$-values (large surfaces). The horizontal line corresponding to a committor value of ½ identifies the members of the transition state ensemble. For the small surfaces (Figure 4), the transition state is mostly composed of configurations with a single molecule remaining in the confined region. It can be seen that even when as few as 3 or 4 molecules remain in the confined space, the majority of trajectories initiated from such configurations return to the liquid state. Figure 5 (large surfaces) shows a different picture. The curve corresponding to $176 \leq N \leq 180$, for which the majority of configurations lead to evaporation, crosses the 50% committor value almost orthogonally (compare with the behavior of the $N = 1$ curve in Figure 4). This indicates that $N$ by itself is not a good order parameter for identifying transition states, a conclusion substantiated in Figure 6. Shown there are three configurations corresponding to $N = 179, 180$ and $190$ (panels a, b, c respectively). The committor probability of the configuration shown in Figure 6(a) is only 4%, even though the number of water molecules in the confined region, 179, is the smallest of the three cases considered. The committor probabilities for the configurations shown in Figures 6(b) and 6(c) are 52 and 83%, respectively. It is clear that the pathway to evaporation involves the formation of a vapor tube of critical diameter (23). The configuration depicted in Figure 6(a), though "farther along" the route towards the vapor phase as measured by $N$, is in reality very far from vaporizing, as it lacks a sufficiently large cavity.

**Conclusions**

The present calculations suggest that there is a narrow range of gaps (~ 5 to 20 Å) between hydrophobic surfaces within which capillary evaporation occurs at rates that may be relevant to biological assembly phenomena. Over the range of gaps (9.0 to 14 Å), surface areas (1 to 9 nm$^2$) and temperatures investigated here ($298 \leq T \leq 398K$), the predominant contribution to the free energy barrier to evaporation comes from line tension. We find that free energy barriers are predominantly enthalpic and increase in proportion to the gap between surfaces at a rate of 4-5 $kT$/Å. We observe a marked increase in the rate of capillary evaporation (on a per unit area basis) upon increasing the size of the hydrophobic surface. Recent simulations have shown that capillary drying is involved in the closing of the pentameric pore in a ligand-gated ion channel (29). The possible relevance of capillary drying to other biophysical phenomena deserves investigation.

FFS is a powerful technique that enables rate calculations spanning more than ten orders of magnitude (e.g., characteristic evaporation times ranging from 6 x 10$^{-10}$ to 17 s for the 1 x 1 nm$^2$ surfaces; see Figure 2 and text). Numerical analysis of the transition state ensemble shows that, for sufficiently large surfaces, the critical nucleus is a gap-spanning cylindrical vapor tube. On smaller surfaces, the transition state ensemble consists largely of configurations containing as little as a single water molecule in the confined space.

**Methods**
**Forward flux sampling.** Consider a system with two locally stable states designated by $A$ (e.g., confined liquid) and $B$ (e.g., confined vapor), which are separated by a free energy barrier much larger than the thermal energy. The goal is to find the thermally-averaged rate at which the system evolves from $A$ to $B$. Consider a property that can distinguish state $A$ from state $B$. For the present problem, it is clear that the number of water molecules in the confined region, $N$, is such a property. For convenience we consider the corresponding intensive property, $\rho$, the average value of which is $\rho_A$ in state $A$ and $\rho_B$ in state $B$ ($\rho_A > \rho_B$). The evolution from *A* to *B* can be described by "interfaces" $\lambda_i$ (i = 0, 1, 2, 3.... ,n) which are collections of configurations with the same value of $\rho$, say $\rho_i$ (31-33). Let $\rho_i > \rho_{i+1}$. State *A* (liquid) is uniquely defined as comprising all configurations with $\rho > \rho_0$ and state *B* (vapor) comprises configurations with $\rho < \rho_n$. We chose $\rho_0$ to be one standard deviation away from the mean liquid density in the confined region. FFS comprises of two steps. In the first step, we calculated the flux from state *A* to the first interface $\lambda_1$ (42, 43). An O(50 nsec) MD simulation was conducted at liquid conditions, and each time the simulation reached $\lambda_1$, the configuration at $\lambda_1$ was stored. Since *A* is the locally stable state, on most occasions, the trajectory reaching $\lambda_1$ returned back to *A*. If on a rare occasion it reached *B*, then the simulation was stopped and restarted from a random initial condition in *A*. The flux to reach $\lambda_1$ from *A* was calculated by dividing the number of $\lambda_1$ crossings that originate from $\lambda_0$ by the total time spent by the MD trajectory within the liquid basin ($\rho > \rho_0$). $\lambda_1$ was chosen so as to ensure that 500-700 independent trajectories from $\lambda_0$ cross $\lambda_1$. Uncorrelated configurations were ensured by storing configurations separated by at least 2 ps. In the second step, the conditional probability of a trajectory starting from $\lambda_i$ and reaching $\lambda_{i+1}$ before reaching $\lambda_0$ [denoted by $P(\lambda_{i+1}|\lambda_i)$] is determined. In order to find $P(\lambda_2|\lambda_1)$, a number of MD trajectories are started from the configurations stored at $\lambda_1$ after velocity randomization, and are propagated until they reach either $\lambda_2$ or $\lambda_0$. $P(\lambda_2|\lambda_1)$ is simply the fraction of trajectories that reach $\lambda_2$ out of all the trajectories started from $\lambda_1$. The configurations at $\lambda_2$ are stored for further propagation to $\lambda_3$ and steps are repeated until the system reaches $\lambda_n$ The rate of the transition from *A* to *B* is then given by (31-33, 42, 43)

$$\text{Rate} = \varphi(\lambda_1|\lambda_0) \prod_i P(\lambda_{i+1}|\lambda_i) \quad (i=1,2,\ldots n-1) \quad (7)$$

where $\varphi(\lambda_1|\lambda_0)$ is the flux of trajectories that leave $\lambda_A$ ($\rho > \rho_0$) and reach $\lambda_1$. Interfaces $\lambda_i$ were chosen to ensure that similar statistics of trajectory crossings are obtained at each interface. From each configuration at $\lambda_i$, 100 trajectories are shot (each with randomized velocities), and the location of $\lambda_{i+1}$ is selected such that $P(\lambda_{i+1}|\lambda_i) \sim 0.01$. Numerical checks were conducted for both the small (1 x 1 nm$^2$) and large (3 x 3 nm$^2$) walls. In the former case, for *d* = 9 Å at 298K evaporation occurred fast enough that it could be computed directly by MD. Comparison of FFS and direct MD rates yielded excellent agreement (1.67 x 10$^9$ vs 1.69 x 10$^9$ nm$^{-2}$ s$^{-1}$, respectively, the latter averaged over 127 evaporation transitions). For the large walls case, the number of interfaces for the case *d* = 14 Å at 334 K was changed from 3 (*N* = 260, 240, 226) to 4 (*N* = 260, 240, 226, 200). The calculated evaporation rates were 1.21 x 10$^5$ and 1.22 x 10$^5$ nm$^{-2}$ s$^{-1}$.

**Transition state ensemble.** The three-step procedure for harvesting the transition state ensemble is as follows. In the first step, an appropriate value was determined for the number of confined water molecules characterizing configurations from which subsequent trajectory "launches" were performed. This appropriate number was determined by starting molecular dynamics runs from states along the various interfaces $\lambda_i$ used in the evaporation rate calculations and identifying an interface from which the probability of reaching the vapor state is significantly less than 1 but non-vanishing. The number of confined water molecules so selected was 7 and 198 for the small and large walls, respectively. At 298K and 1 bar, the corresponding probabilities of reaching the vapor phase were 0.026 and 0.015, respectively. In the second step, molecular dynamics runs were launched from these starting configurations (i.e., from configurations with 7 and 198 water molecules confined between the small and large walls, respectively); a subset of these reached successive interfaces on the way to the vapor phase, and these configurations were saved. For the small wall simulations, $O(10^4)$ runs were launched from $N = 7$, 150 of which reached $N = 1$. Another $O(10^4)$ runs were launched form $N = 7$, of which 400 reached $N = 5$. This yielded $O(10^2)$ configurations in each of the milestones $N = 1, 2, 3, 4$ and 5. Similar calculations for the large walls case yielded $O(10^2)$ configurations in each of the three milestone ranges $176 \leq N \leq 180$, $181 \leq N \leq 185$, and $186 \leq N \leq 190$, the grouping being necessary because of the much larger number of molecules. In the third step, 100 trajectories were launched starting from each of the candidate configurations (i.e., 100 trajectories starting from each of the $O(10^2)$ $N=1$ configurations, 100 from each of the $O(10^2)$ $N = 2$, etc.).

**Molecular dynamics.** Mimicking the arrangement of carbon atoms in graphene sheets, the hydrophobic walls were represented by a rigid, hexagonal lattice of Lennard Jones (LJ) atoms with a lattice constant of 1.4 Å. The walls were kept fixed, parallel to each other, separated by a distance $d$, and symmetrically-located with respect to the center of the simulation box. The SPC/E water model was used throughout (44). The LJ parameters for water-wall interaction were taken as $\varepsilon_{O\text{-}W} = 0.0289$ kcal/mol and $\sigma_{O\text{-}W} = 3.283$ Å (18). MD simulations were conducted in the isothermal-isobaric (NPT) ensemble at 298 K and 1 bar in a periodic simulation box, using a Nose'-Hoover thermostat and barostat (45, 46). All simulations were performed using the LAMMPS MD package (47). The number of simulated water molecules was 2329 for the small wall system and 4685 for the large wall system. The Particle Particle Particle Mesh (PPPM) Ewald method was used to compute long-range corrections to electrostatic interactions (48, 49). The $k$-space vector was taken to be 0.295 Å$^{-1}$, and calculations were performed on a 25 x 36 x 36 grid, with RMS precision of 6 x 10$^{-5}$, the standard PPPM Ewald parameters in LAMMPS.

**Acknowledgements**
The financial support of Unilever U.K. Central Resources is gratefully acknowledged. Computations were performed at the Terascale Infrastructure for Groundbreaking Research in Engineering and Science facility at Princeton University. We are grateful to Alenka Luzar and Amish Patel for insightful comments on this work.

**Figure Captions**

**Figure 1.** Schematic of evaporation rate calculation. Two $L \times L$ hydrophobic surfaces (green atoms), separated by a gap $d$, are immersed in 2329 ($L$ = 1 nm) or 4685 ($L$ = 3 nm) water molecules, at atmospheric pressure. Forward flux sampling simulations (31-33) are carried out to compute the rate of capillary evaporation in the confined region of width $d$, for a range of values of $d$, $L$ and temperature.

**Figure 2.** Calculated evaporation rates. Dependence of the evaporation rate on the gap between 1 x 1 nm$^2$ hydrophobic surfaces, at 298K. The inset shows, for the same surfaces, Arrhenius plots of the evaporation rate for two values of the gap, corresponding to calculations at $T$ = 298, 348 and 398K.

**Figure 3.** Calculated evaporation rates. Dependence of the evaporation rate on the gap between 3 x 3 nm$^2$ hydrophobic surfaces, at 298K.

**Figure 4.** Identification of the transition state ensemble. Each curve gives the probability, computed over 100 runs launched from a given configuration after randomizing the velocities, that such runs will reach the vapor state (no water molecules in the confined region) without first returning to the liquid state. This probability is plotted as a function of configuration number, with configurations ranked in order of increasing committor probability. All of the configurations along a given line have the same number of confined water molecules ($N$ = 0, 1, 2, 3 or 4). Conditions are $d$ = 9.8 Å, $T$ = 298K, $P$ = 1 bar, $L$ = 1 nm. The transition state ensemble corresponds to those configurations with equal probability of reaching the vapor state or of returning to the liquid state (dashed line). Along each line, the number of configurations has been normalized so as to lie between 1 and 100. For example, if there are $m \neq 100$ configurations with $N$ = 1, their number has been scaled by $100/m$.

**Figure 5.** Identification of the transition state ensemble. Same as Figure 4, but for $d$ = 12 Å, $T$ = 298K, $P$ = 1 bar, $L$ = 3 nm. Because of the larger number of confined water molecules compared to the $L$ = 1 case (Figure 4), these have been combined into groups for ease of representation. Thus, each curve corresponds to a range of $N$-values.

**Figure 6.** Selected configurations intermediate between the confined liquid and vapor states, for $d$ = 12 Å, $T$ = 298K, $P$ = 1 bar, and $L$ = 3 nm. The hydrophobic surfaces, whose boundary is traced by the yellow line, have been removed for ease of visualization. The view is along the direction perpendicular to the surfaces. The number of confined water molecules and committor probability for these configurations are (179, 0.04), (180, 0.52) and (190, 0.83) for panels (a), (b) and (c) respectively. Gap-spanning vapor tubes are clearly visible in (b) and (c).

**Table Captions**

**Table 1:** Comparison of free energy barriers for evaporation between small (1 x 1 nm$^2$) and large ( 3 x 3 nm$^2$) surfaces, calculated directly from computed evaporation rates at 298K, and by assuming linear dependence of the barrier on the gap size.

**Table 2:** Comparison of mean density and compressibility of water at 298K and 1 bar confined between small (1 x 1 nm$^2$) and large (3 x 3 nm$^2$) surfaces

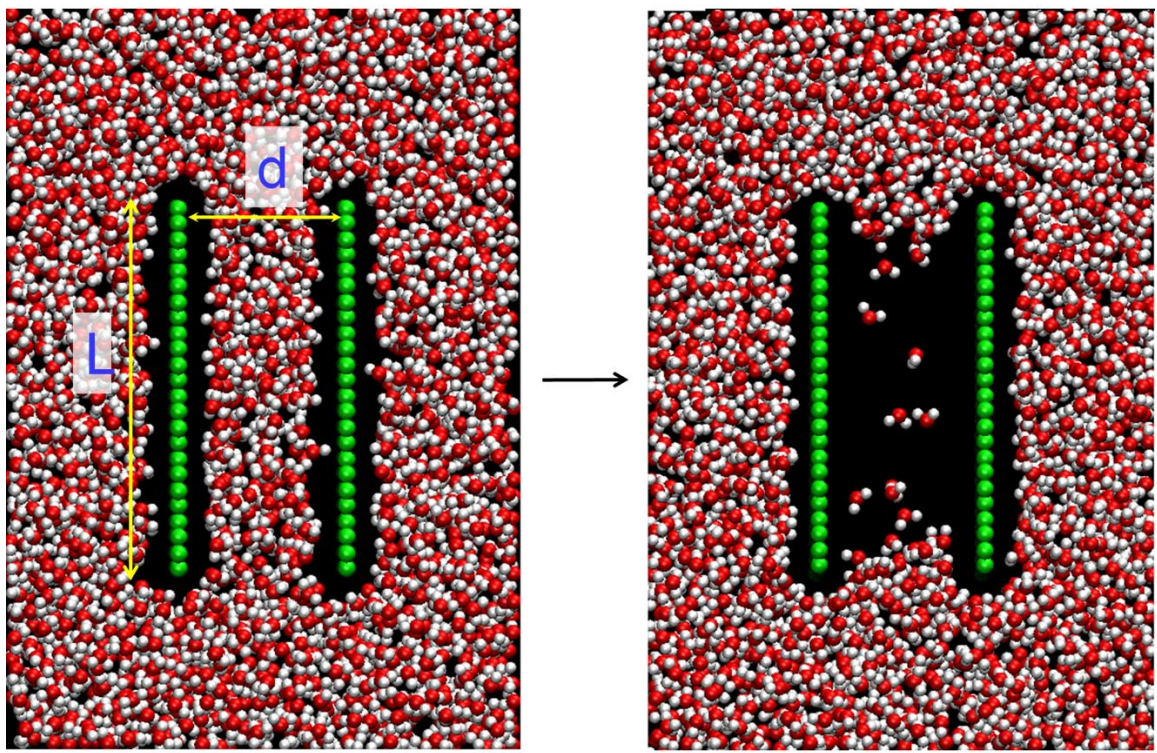

**Figure 1**

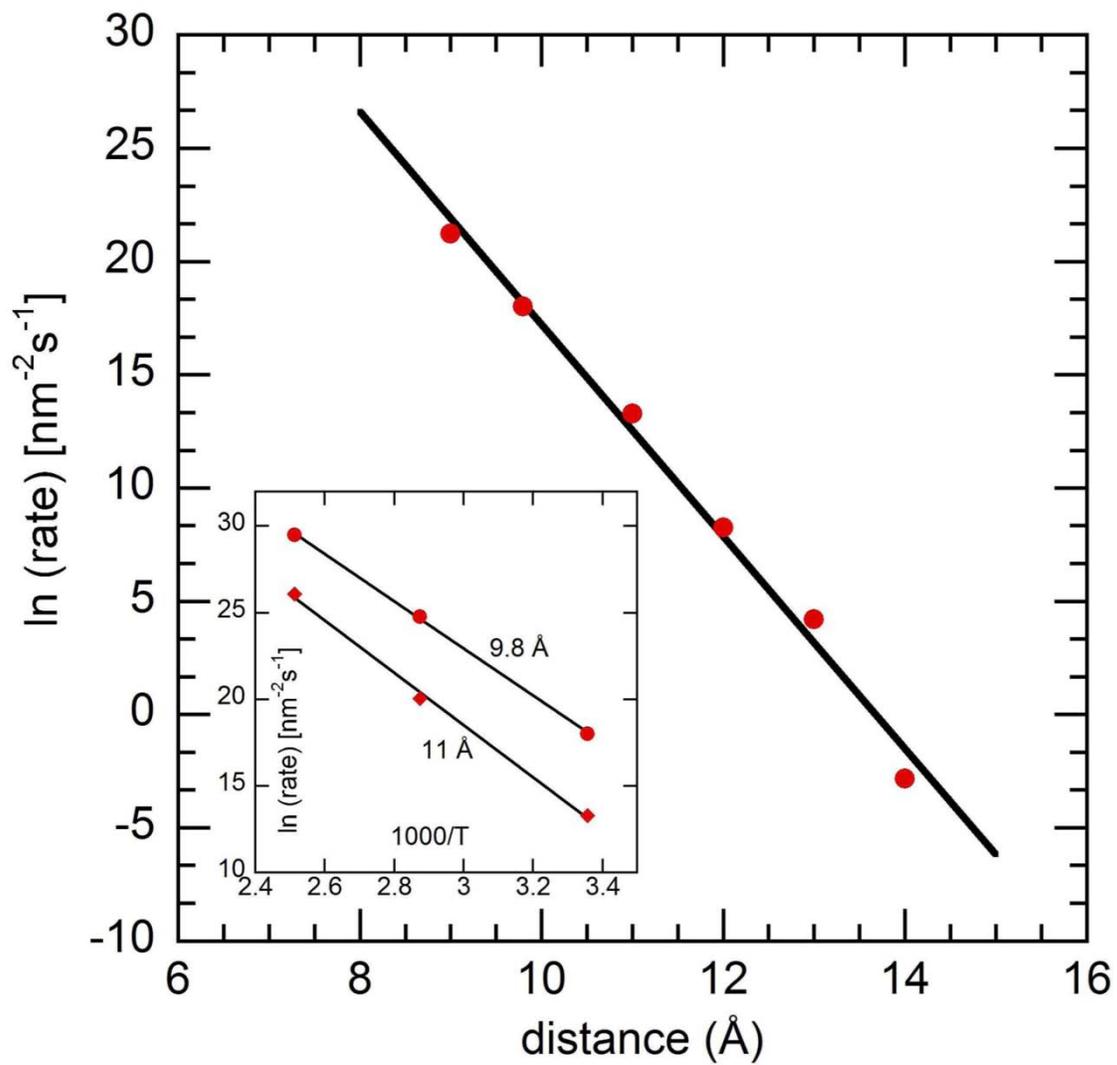

**Figure 2**

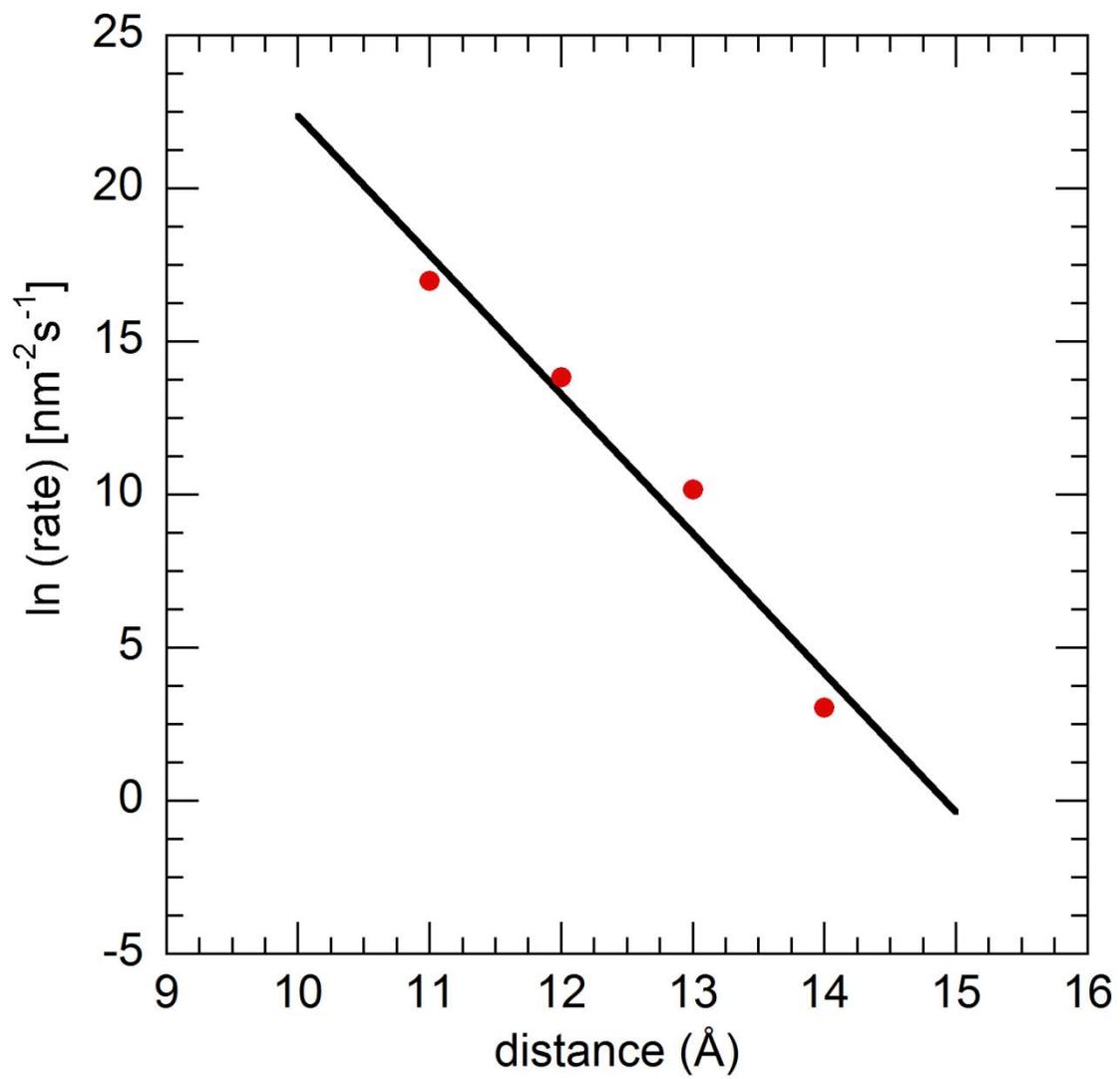

**Figure 3**

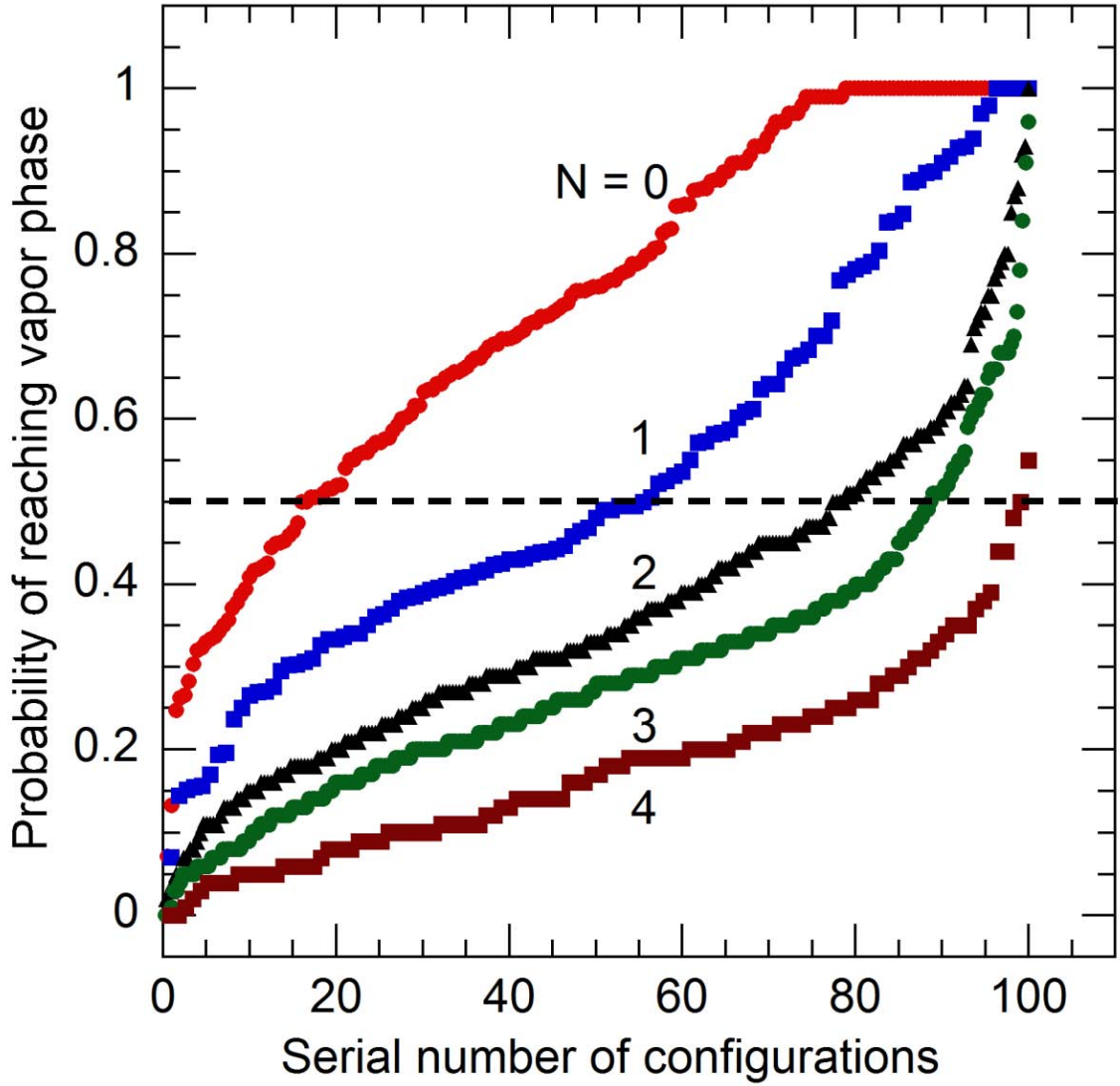

**Figure 4**

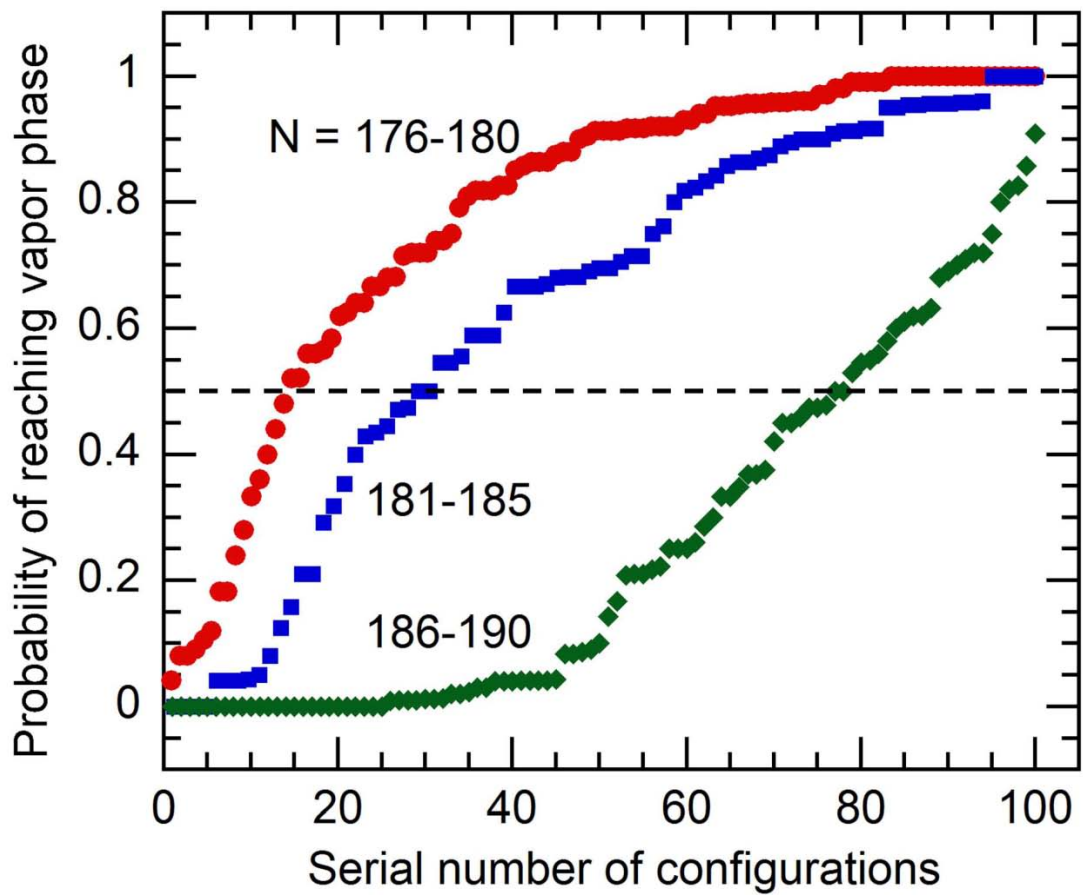

**Figure 5**

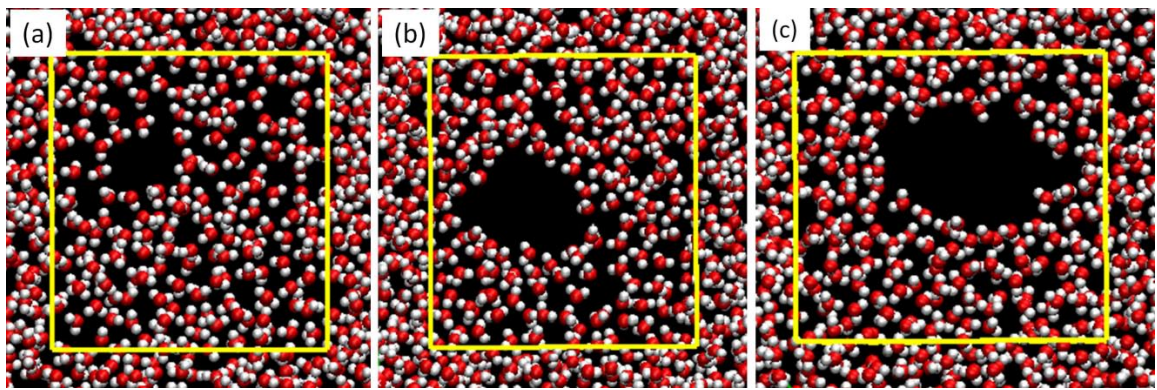

**Figure 6**

**Table 1**

| Gap (Å) | ΔG/kT (small surface) (Equation 1*) | ΔG/kT (small surface) (Linear scaling†) | ΔG/kT (large surface) (Equation 1*) | ΔG/kT (large surface) (Linear scaling†) |
|---|---|---|---|---|
| 9.0 | 42.5 | 42.0 | — | — |
| 9.8 | 45.7 | 45.7 | — | — |
| 11 | 50.4 | 51.3 | 57.8 | 55.8 |
| 12 | 55.5 | 55.9 | 60.9 | 60.9 |
| 13 | 59.5 | 60.6 | 64.5 | 65.9 |
| 14 | 66.5 | 65.2 | 71.7 | 71.0 |

* Free energy barriers obtained from evaporation rate calculations ($\Delta G/kT = \ln C - \ln j$), with pre-exponential factor obtained from as explained in the Supporting Information.
† Free energy barriers calculated assuming linear dependence of $\Delta G$ on $d$, $\Delta G(d) = \Delta G$ (9.8 Å) x $d$/9.8 for small surfaces, and $\Delta G(d) = \Delta G$ (12 Å) x $d$/12 for large surfaces.

**Table 2**

| Gap (Å) | Density[‡] (g/cc) | | $10^5$ x Compressibility[‡] (1/bar) | |
|---|---|---|---|---|
| | Small surface | Large surface | Small surface | Large surface |
| **11** | 1.477 (0.595) | 1.384 (0.558) | 13.5 (33.6) | 18.3 (45.3) |
| **12** | 1.369 (0.620) | 1.268 (0.574) | 13.6 (30.2) | 16.5 (36.4) |
| **13** | 1.332 (0.659) | 1.235 (0.611) | 14.4 (29.2) | 15.3 (30.9) |
| **14** | 1.302 (0.691) | 1.209 (0.642) | 13.8 (26.0) | 14.2 (26.7) |

[‡] The width of the confined region was calculated as $d - 2\sigma_{O-W}$ (numbers without parenthesis) or as $d$ (numbers in parenthesis).

# Supporting Information

## Derivation of Equation 2

Consider the formation of a cylindrical vapor tube of radius $r$, spanning the gap $d$ between two circular solid surfaces of radius $L$ immersed in a liquid (Figure S1). The resulting free energy change within the open system boundary shown in Figure S1 is given by

$$\Delta\Omega = -P_v \pi r^2 d - P_l \pi (L^2 - r^2) d + \gamma_{vl} 2\pi r d + 2\pi r^2 \gamma_{vs} + 2\pi (L^2 - r^2) \gamma_{ls} + 4\pi r \lambda + P_l \pi L^2 d - 2\pi L^2 \gamma_{ls} \tag{S.1}$$

where $P_v$ and $P_l$ denote the pressures inside and outside of the vapor tube, respectively; $\gamma_{vl}$, $\gamma_{vs}$ and $\gamma_{ls}$ denote the vapor-liquid, vapor-solid and liquid-solid interfacial tensions, respectively; and $\lambda$ is the line tension. We consider hard walls (contact angle $\pi$), and invoke the condition of mechanical equilibrium to relate $P_l$ and $P_v$,

$$\gamma_{ls} - \gamma_{vl} = \gamma_{vs} \tag{S.2}$$

$$P_v - P_l = \frac{\gamma_{vl}}{r} \tag{S.3}$$

Substitution of (S.2.) and (S.3) into (S.1) yields Equation (2), with $\gamma$ denoting the vapor-liquid surface tension.

## Modification of Equation 2 for the case of length scale-dependent surface tension

We investigate whether the linear relationship between the free energy barrier and the gap between hydrophobic surfaces can be explained by invoking the dependence of surface tension on the size of the vapor cylinder. To this end, we write (1)

$$\gamma_{vl} = \gamma^\infty \left(1 - \frac{2\delta}{r}\right) \tag{S.4}$$

where $\delta$ is the Tolman length. Substituting in Equation (2), setting $\lambda = 0$, and differentiating with respect to $r$ yields the free energy maximum condition analogous to Equation (3),

$$r^* = \frac{d}{4} + \delta \tag{S.5}$$

in correspondence to which the free energy barrier is given by

$$\Delta\Omega = \frac{\pi\gamma^{\infty}d^2}{8} - \pi\gamma^{\infty}\delta d + 2\pi\gamma^{\infty}\delta^2 \tag{S.6}$$

Using typical values for water [$d \sim 1$nm, $\delta \sim 1$ Å (2)], the relative magnitude of the three terms on the right hand side of Equation (S.6) is $\sim$ 1: 0.8: 0.08. This implies that for the characteristic dimensions ($d$) and substance ($\delta$) considered here, the size dependence of the surface tension gives rise to a contribution to the free energy barrier that is linear in $d$; this term, however, is at best comparable to the quadratic contribution. In contrast, as shown in the body of the paper, allowing for line tension effects gives rise to terms that are an order of magnitude larger than the quadratic contribution. Thus, the observed linearity of the free energy barrier with respect to $d$ cannot be explained in terms of the size-dependence of the surface tension.

**Free energy barrier calculations**
*Small surfaces* According to Eq. (1), an Arrhenius plot of ln*j* versus *1/T* at fixed *d* should yield the slope, $-\Delta H/k$, and the intercept, $\ln C' = \ln C + \Delta S/k$. For the 1 x 1 nm$^2$ surfaces, evaporation rates were calculated at 298, 348 and 398K for *d*=9.8 and 11Å (Figure 2, inset), and ln*C'* and $\Delta H(d)$ were determined using an ordinary least square fit to the ln*j* versus *1/T* data for both *d*=9.8 and 11Å. The regressed value of ln*C'* [with *C'* in nm$^{-2}$s$^{-1}$] was found to be 63.8 for both *d*=9.8 Å and *d*=11 Å. The distance-independent estimate of *C'* implies that either $\Delta S(d)$ is independent of *d* or is small. Using *C'*, the values of $\Delta H$ were determined from the calculated evaporation rates at 298K for different *d* (see Eq. 1). The calculated values of $\Delta H$ were found to scale linearly with *d*. Since the entropic contribution is small or independent of *d*, $\Delta G$ should also scale approximately linearly with *d*. Writing $\Delta G(d) = A + Bd$, where *A* and *B* are unknown constants, and substituting in Eq. (1), we get

$$j = C \exp(-A/kT) \exp(-Bd/kT) \tag{S.7}$$

According to Eq.(S.7), a plot of ln*j* versus *d* at fixed *T* should give $-B/kT$ as the slope and $\ln C - A/kT$ as the intercept. To estimate the values of *C*, *A* and *B*, we performed an ordinary least square fit to Eq. (S.7) using the evaporation rate data for the 1x1 nm$^2$ surfaces for different *T* and *d* (see Table S1). The values of ln*C*, *A/k* and *B/k* were found to be 63.7, -312.7 K and 1413.1K/Å respectively. $\Delta S/k$, from Eq. (1), is equal to $\ln(C'/C) = 0.1$. Hence, $T\Delta S/\Delta H$ is O(10$^{-3}$). Therefore, $\Delta G \approx \Delta H$.

*Large surfaces* For the 3 x 3 nm$^2$ surfaces, evaporation rates were calculated at 298, 334 and 360K for *d*=14 Å and at 334, 364 and 390K for *d*=16Å (see Table S1). The regressed values of ln*C'* [*C'* in nm$^{-2}$s$^{-1}$] using Eq. (1) were found to be 77.2 for *d*=14 Å and 82.7 for *d*=16 Å. The estimated value of $\Delta H$ for *d*=14Å was 60.2 *kT*. The variation in the value of ln*C'* is small in comparison to value of *ΔH/kT*. Taking ln*C'* as 77.2, the values of *ΔH* can be estimated from the calculated evaporation rates at 298K. For the 3 x 3 nm² surfaces also, to a very good approximation, *ΔH* values were found to scale linearly with *d*. Using the evaporation rate data for the 3x3 nm² surfaces for different *d* and *T* (Table S1), an ordinary least square fit to Eq. (S.7) yielded values for ln*C*, *A/k* and *B/k* = 74.7, 484.0K and 1478.7K/Å respectively. *ΔS/k* was found to vary between ln($C'/C$) = 2.5 (for d = 14 Å) to 8.0 (for 16 Å). Hence, *TΔS/ΔH* is O($<10^{-1}$). Thus, for large surfaces also, *ΔG* ≈ *ΔH*.

**Scaling of the Free Energy Barrier**

Having obtained *C* as explained above, *ΔG(d)/kT* was calculated from the computed evaporation rates as a function of *d* at 298K, using Eq. (1). Figures S2 (1 x 1 nm² surfaces) and S3 (3 x 3 nm² surfaces) compare the free energy barrier for evaporation, *ΔG/kT*, computed directly from Eq. 1, to the quadratic scaling (*ΔG/kT* ~ *d²*), and to the scalings *ΔG/kT* ~ *(d-2l)* and *ΔG/kT* ~ *(d-2l)²*, where *l* is the width of the vapor layer. The width of the vapor layer is the closest distance to the wall that a water molecule in the confined region is able to reach during the MD simulation. It was found to be ca. 2.4 Å (Figure S4) across the range of conditions investigated in this work. As seen from Figures S2 and S3, the linear scaling in *d* shows good agreement with the calculated values of *ΔG* for both the small and the large surfaces, whereas other scaling relations [~ *d²*, *(d-2l)* and *(d-2l)²*] fail.

**Effect of Line Tension on Critical Gap**

Consider two parallel *L* x *L* hydrophobic surfaces immersed in water, separated by a gap *d*. Comparing the free energy of the liquid and vapor phases in the *L* x *L* x *d* region between surfaces, we can determine the critical distance, $d_c$, below which the confined liquid becomes metastable with respect to the confined vapor (3-5). Taking into account line tension, the free energies in the confined region are given by

$$\Omega_l = -P_l L^2 d + 2\gamma_{ls} L^2 \tag{S.8}$$

$$\Omega_v = -P_v L^2 d + 2\gamma_{vs} L^2 + 4\gamma_{vl} L d + 8\lambda L \tag{S.9}$$

where the symbols have already been introduced in Eq. (S.1). $d_c$ is then calculated from the condition $\Omega_v = \Omega_l$, together with (S.2) (it is assumed here that the contact angle of water on the hydrophobic surfaces is 180°):

$$d_c = \frac{2\gamma - 8\lambda/L}{\Delta p \left(1 + 4\gamma/L\Delta p\right)} \tag{S.10}$$

where $\gamma$ denotes the vapor-liquid surface tension, and $\Delta p = P_l - P_v$. For water at ambient conditions and $L \sim O(1 \text{ nm})$, $4\gamma/L\Delta p \sim 10^3$, and (S.10) reduces to

$$d_c = \frac{L}{2}\left(1 - \frac{4\lambda}{\gamma L}\right) \tag{S.11}$$

For water at ambient temperature, and using an order of magnitude estimate for $\lambda$ (6, 7), one obtains $1 < 4\lambda/\gamma L < 10$. Assuming the validity of macroscopic reasoning at these length scales, Eq. (S.11) shows that line tension can either reduce or increase $d_c$, depending on the sign of $\lambda$.

**Figure Captions**

**Figure S1.** Schematic diagram showing boundaries and relevant dimensions for the calculation of the free energy barrier to the formation of a vapor tube.

**Figure S2.** Comparison of different possible scaling relations of the free energy barrier, $\Delta G/kT$, with the distance, $d$ between the small (1 x 1 nm$^2$) hydrophobic surfaces. The scaling with $d$ shows excellent agreement with the calculated $\Delta G/kT$. The line is a guide to the eye.

**Figure S3.** Comparison of different possible scaling relations of the free energy barrier, $\Delta G/kT$, with the distance, $d$ between the large (3 x 3 nm$^2$) hydrophobic walls. The scaling with $d$ shows excellent agreement with the calculated $\Delta G/kT$. The line is a guide to the eye.

**Figure S4.** Density profile of water for small (1 x 1 nm$^2$) surfaces with $d$ = 14 Å with the vapor layer identified. The LJ centers corresponding to the wall atoms are located at -7.0 Å and +7.0 Å. The first non-zero value for water's density profile defines the width of the vapor layer, $l$.

**Table Captions**

**Table S1:** Evaporation rates obtained from Forward Flux Sampling calculations performed in this work for various values of temperature, $T$, gap between hydrophobic surfaces, $d$, and linear dimension of the hydrophobic surface, $L$.

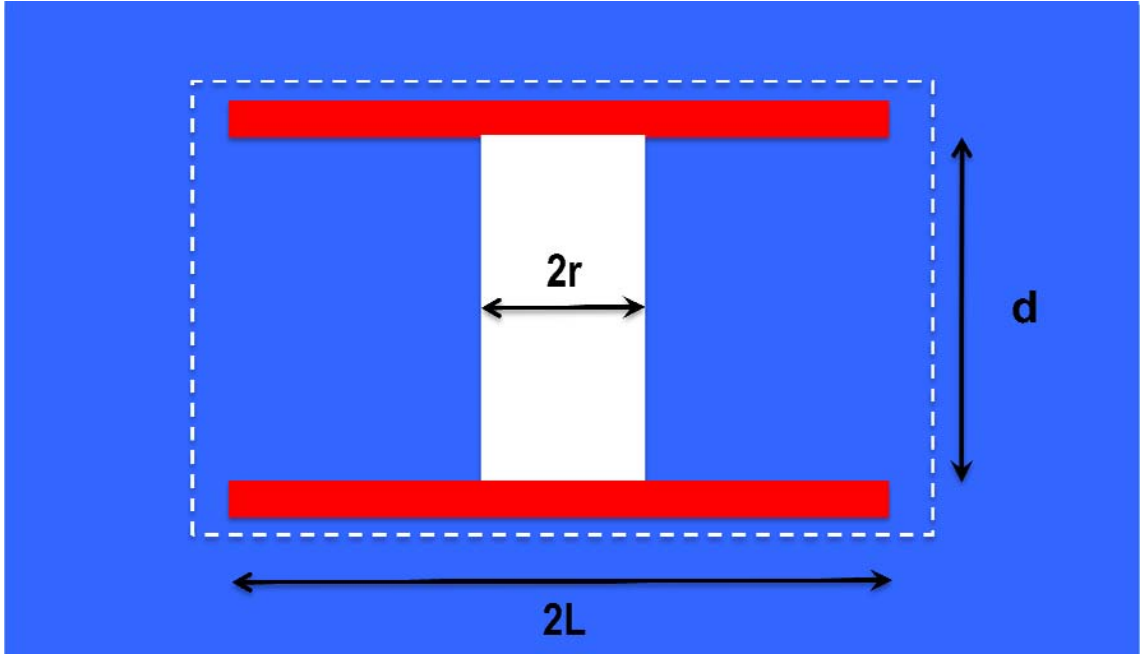

**Figure S1**

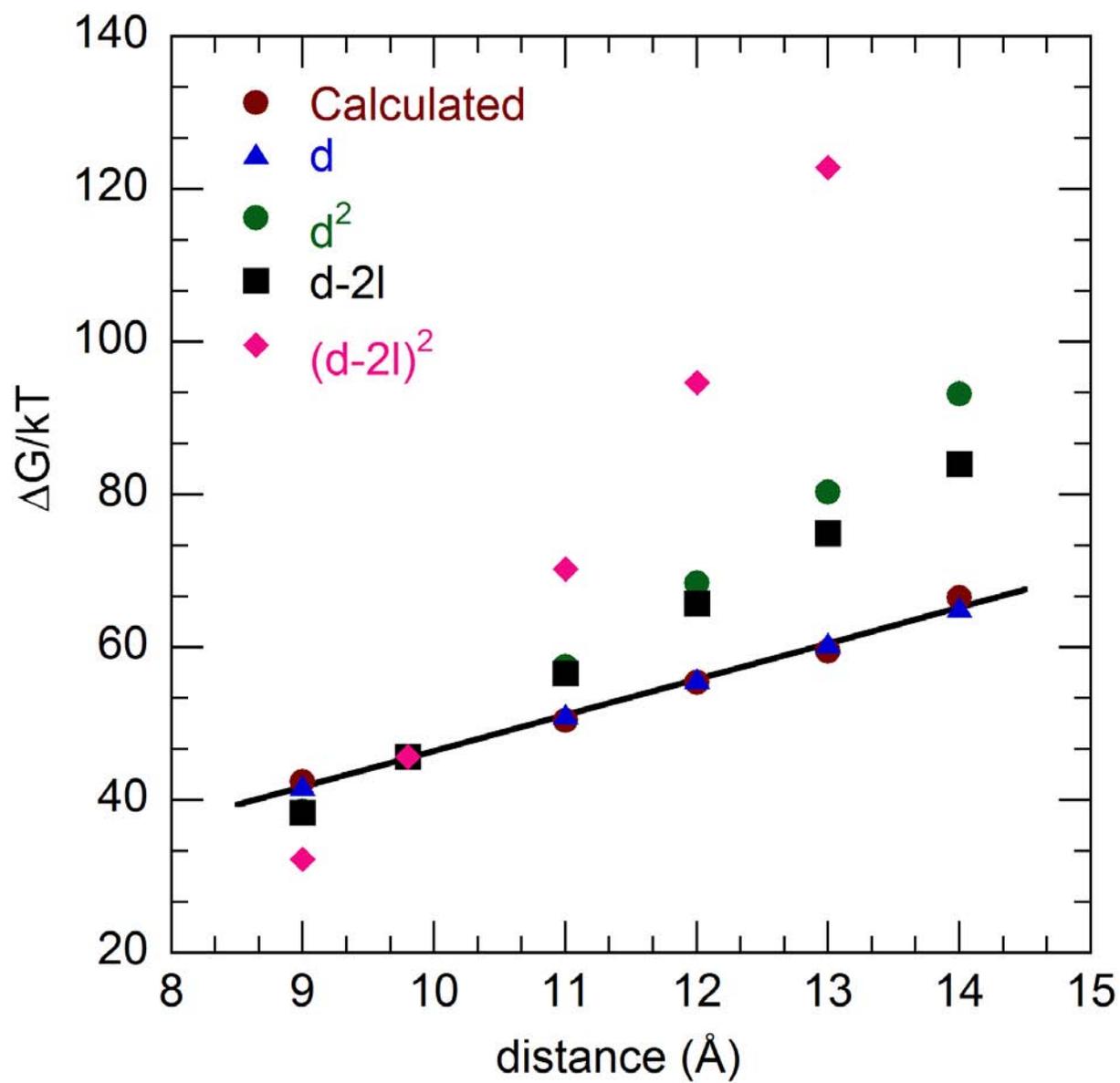

**Figure S2**

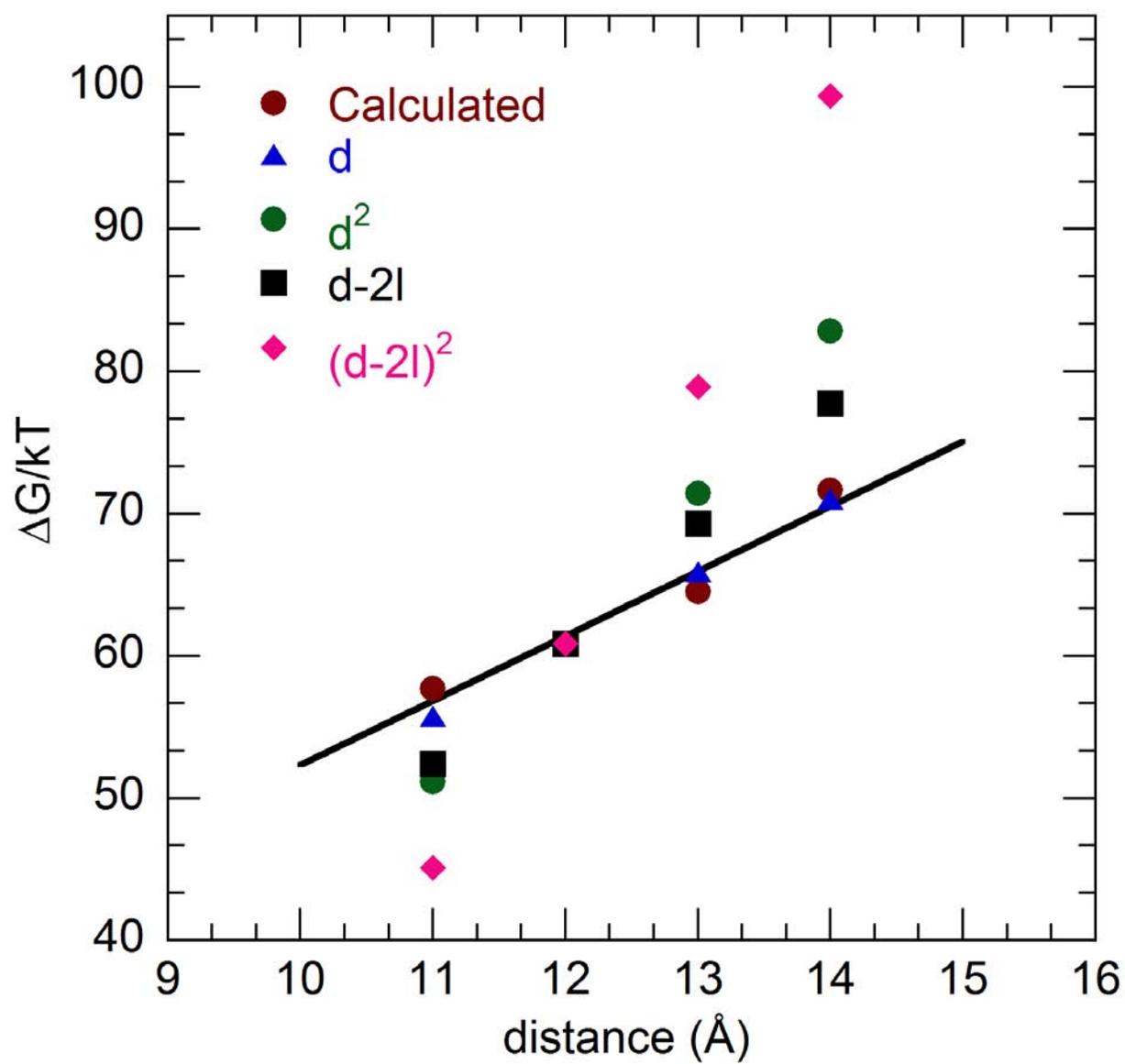

**Figure S3**

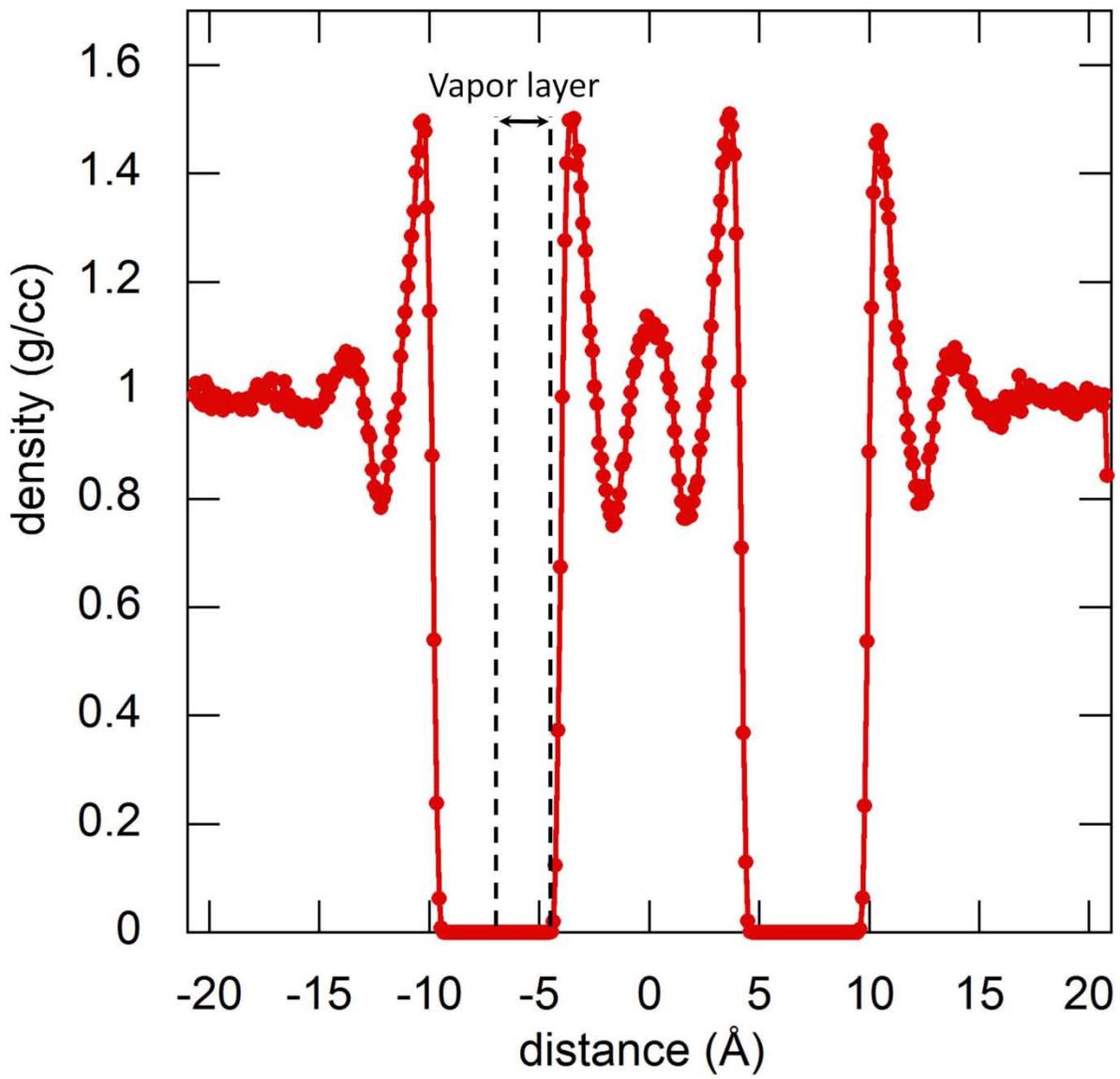

**Figure S4**

**Table S1**

| T (K) | d (Å) | L (nm) | ln(j) (j in nm$^{-2}$s$^{-1}$) |
|---|---|---|---|
| 298 | 9.0 | 1 | 21.2 |
| 298 | 9.8 | 1 | 18.0 |
| 298 | 11 | 1 | 13.3 |
| 298 | 12 | 1 | 8.2 |
| 298 | 13 | 1 | 4.2 |
| 298 | 14 | 1 | -2.8 |
| 348 | 9.8 | 1 | 24.8 |
| 348 | 11 | 1 | 20.1 |
| 398 | 9.8 | 1 | 29.5 |
| 398 | 11 | 1 | 26.1 |
| 298 | 11 | 3 | 17.0 |
| 298 | 12 | 3 | 13.8 |
| 298 | 13 | 3 | 10.2 |
| 298 | 14 | 3 | 3.0 |
| 334 | 13 | 3 | 15.9 |
| 334 | 14 | 3 | 11.7 |
| 334 | 15 | 3 | 8.1 |
| 334 | 16 | 3 | 0.7 |
| 360 | 14 | 3 | 15.6 |
| 364 | 16 | 3 | 7.5 |
| 390 | 16 | 3 | 12.5 |